\documentclass[letterpaper,11pt]{article}

\usepackage[utf8]{inputenc}
\usepackage[T1]{fontenc}
\usepackage{cfr-lm}

\usepackage{stmaryrd}
\usepackage{cite}
\usepackage[dvips]{graphicx}
\usepackage{graphpap}
\usepackage{ifthen}
\usepackage{enumerate}
\usepackage{amssymb}
\usepackage{mathrsfs}
\usepackage[errorshow]{tracefnt}
\usepackage{fancybox}
\usepackage{amsfonts}
\usepackage{amsmath}
\usepackage{amssymb}
\usepackage{multirow}
\usepackage{array}
\usepackage{mathtools}
\usepackage{soul}
\usepackage{pict2e}
\usepackage{bbold}
\DeclareMathAlphabet{\mathpzc}{OT1}{pzc}{m}{it}

\usepackage{lscape}
\usepackage{xypic}
\usepackage[dvipsnames]{xcolor}
\usepackage[
color,matrix,arrow]{xy}
\usepackage{setspace}

\usepackage{epsf}

\usepackage{enumitem}

\usepackage{relsize}
\usepackage{exscale}

\usepackage{graphicx}

\renewcommand{\geq}{\geqslant}

\makeatletter
\newcommand{\thickhline}{%
    \noalign {\ifnum 0=`}\fi \hrule height 1pt
    \futurelet \reserved@a \@xhline
}
\newcolumntype{'}{@{\hskip\tabcolsep\vrule width 1pt\hskip\tabcolsep}}
\makeatother
\newcolumntype{"}{@{\hskip\tabcolsep\vrule width 1.5pt\hskip\tabcolsep}}
\makeatother

\newcommand{\scr}{\mathscr}

\def\ie{{\it i.e.}}

\def\eg{{\it e.g.}}
\def\cf{{\it cf.}}
\newcommand\BB{{\scr B}}

\def\small#1{{\hbox{$#1$}}}
\def\fraction#1{\small{1\over#1}}
\def\fr{\fraction}

\def\boxit#1{\vbox{\hrule\hbox{\vrule\kern3pt
             \vbox{\kern3pt#1\kern3pt}\kern3pt\vrule}\hrule}}

\newcommand{\Red}[1]{{\color{red} #1}}

\newcommand{\beq}{\begin{equation}}
\newcommand{\beqn}{\begin{equation*}}
\newcommand{\eeq}{\end{equation}}
\newcommand{\eeqn}{\end{equation*}}
\newcommand{\beqa}{\begin{eqnarray}}
\newcommand{\beqan}{\begin{eqnarray*}}
\newcommand{\eeqa}{\end{eqnarray}}
\newcommand{\eeqan}{\end{eqnarray*}}
\newcommand{\bdm}{\begin{displaymath}}
\newcommand{\edm}{\end{displaymath}}

\newcommand{\ba}{\begin{array}}
\newcommand{\ea}{\end{array}}

\newcommand\nn{\nonumber}

\newcommand\benu{\begin{enumerate}}
\newcommand\eenu{\end{enumerate}}
\newcommand\bit{\begin{itemize}}
\newcommand\eit{\end{itemize}}

\def\End{\mathrm{End\,}}

\def\der'{\mathfrak{der}'\,}
\def\der{\mathfrak{der}\,}
\def\str'{\mathfrak{str}'\,}
\def\str{\mathfrak{str}\,}

\newcommand{\dd}{{\mathsf{d}}}
\newcommand{\KK}{{\mathsf{K}}}

\newcommand{\dlb}{\ensuremath{[\![}}
\newcommand{\drb}{\ensuremath{]\!]}}

\newcommand{\ad}{\text{ad}\,}

\def\fg{{\mathfrak g}}
\def\fgr{{\mathfrak g}}
\def\fgrplus{{\mathfrak g}^+}

\def\sB{{\scr B}}

\def\sh{\sharp}
\def\fl{\flat}

\def\*{\partial}

\def\tk{\widetilde k}
\def\tR{\widetilde R}

\def\id{\mathbb{1}}    

\def\shift#1#2{\underset
  {\scriptscriptstyle\Red{#1}}{#2{}^{\mathstrut}_{\mathstrut}}}

\def\adj{\hbox{\bf adj}}

\def\RR{{\mathbb R}}

\def\LL{{\scr L}}
\def\MM{{\scr M}}

\RequirePackage{hyperref}

\numberwithin{equation}{section}

\begin{document}

\frenchspacing

\null\vspace{-18mm}

\includegraphics[width=4cm]{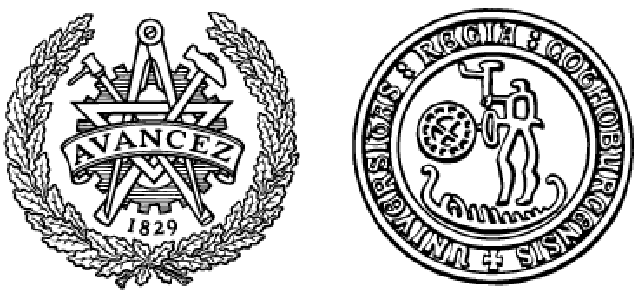}

\vspace{-17mm}
{\flushright Gothenburg preprint \\ August, 2019\\}

\vspace{4mm}

\hrule

\vspace{12mm}

\pagestyle{empty}

\begin{center}
  {\Large \bf \sc Tensor hierarchy algebras and extended geometry II:}
  \\[3mm]
  {\large \bf \sc Gauge structure and dynamics}
    \\[8mm]

{\large
Martin Cederwall${}^1$ and Jakob Palmkvist${}^{1,2}$}

\vspace{8mm}
       ${}^1${\it Division for Theoretical Physics, Department of Physics,\\
         Chalmers University of Technology,
 SE-412 96 Gothenburg, Sweden}

\vspace{5mm}
       ${}^2${\it Division for Algebra and Geometry, Department of
         Mathematical Sciences,\\
         Gothenburg University and Chalmers University of Technology,\\
 SE-412 96 Gothenburg, Sweden}

\end{center}

\vfill

\begin{quote}

\textbf{Abstract:} The recent investigation of the gauge structure of
extended geometry is generalised to situations when ancillary
transformations appear in the commutator of two generalised
diffeomorphisms. The relevant underlying algebraic structure turns out
to be a tensor hierarchy algebra rather than a Borcherds
superalgebra.
This tensor hierarchy algebra is a non-contragredient superalgebra,
generically infinite-dimensional, which is a double extension
of the structure algebra of the extended geometry.
We use it to perform a (partial)
analysis of the gauge structure in terms of an $L_\infty$ algebra for
extended geometries based on finite-dimensional structure groups.
An invariant pseudo-action is also given in these cases.
We comment on the continuation to infinite-dimensional structure groups.
An accompanying paper \cite{CederwallPalmkvistTHAI}
deals with the mathematical construction of the
tensor hierarchy algebras.

\end{quote} 
\vfill

\hrule

\noindent{\tiny email:
  martin.cederwall@chalmers.se, jakob.palmkvist@chalmers.se}

\newpage

\tableofcontents

\pagestyle{plain}

\section{Introduction}

Extended geometry \cite{Cederwall:2017fjm,Cederwall:2018aab}
has recently emerged as a unified framework, with
double and exceptional geometry as special cases where string theory
dualities are geometrised.
It has been shown to have close relations to
certain classes of Lie superalgebras
\cite{Palmkvist:2015dea,Cederwall:2015oua,Cederwall:2017fjm,Cederwall:2018aab}.
The purpose of the present paper is to demonstrate how indeed the
recently invented {\it tensor hierarchy algebras} (THA's)
\cite{Palmkvist:2013vya,Carbone:2018xqq}
are to be seen as the algebraic structure
responsible for and underlying extended geometry, in its most general
setting. 
Besides unifying double geometry
\cite{Tseytlin:1990va,Siegel:1993xq,Siegel:1993bj,Hitchin:2010qz,Hull:2004in,Hull:2006va,Hull:2009mi,Hohm:2010jy,Hohm:2010pp,Jeon:2012hp,Park:2013mpa,Berman:2014jba,Cederwall:2014kxa,Cederwall:2014opa,Cederwall:2016ukd}
and exceptional geometry
\cite{Hull:2007zu,Pacheco:2008ps,Hillmann:2009pp,Berman:2010is,Berman:2011pe,Coimbra:2011ky,Coimbra:2012af,Berman:2012vc,Park:2013gaj,Cederwall:2013naa,Cederwall:2013oaa,Aldazabal:2013mya,Hohm:2013pua,Blair:2013gqa,Hohm:2013vpa,Hohm:2013uia,Hohm:2014fxa,Cederwall:2015ica,Butter:2018bkl,Bossard:2017aae,Bossard:2018utw,Bossard:2019ksx},
one of the advantages of the framework of extended geometry is that it opens a window to situations with
infinite-dimensional structure groups
\cite{Bossard:2017aae,Bossard:2018utw,Bossard:2019ksx}. Eventually, we
would like to establish contact with the
$E_{10}$ \cite{Damour:2002cu}
and $E_{11}$ \cite{West:2001as} proposals.
In the present paper, we will however
limit our attention to finite-dimensional structure groups.

We consider extended geometry based on a Kac--Moody algebra $\fgr$ of rank $r$
and a dominant integral weight $\lambda$, as defined in
ref. \cite{Cederwall:2017fjm}. 
Here, we assume $\fgr$ to be finite-dimensional and simply laced (or
at least that the Dynkin labels $\lambda_i$ corresponding to short roots
$\alpha_i$ vanish).
This still includes ordinary, double and (up to $r=8$) exceptional
geometry, as well as many other, more ``exotic'' extended geometries. 
We are particularly interested in the cases where 
the highest root $\theta$ of $\fgr$ satisfies $(\lambda,\theta) >1$,
which implies that $\fgrplus$ is infinite-dimensional.
As shown in ref. \cite{Cederwall:2017fjm}, these are exactly the cases
where the generalised diffeomorphisms do not close into themselves, 
but only up to ``ancillary'' 
$\fgr$ transformations
\cite{Hohm:2013jma,Hohm:2014fxa,Cederwall:2015ica,Bossard:2017aae}.
For this reason, they were not included in
the analysis in ref. \cite{Cederwall:2018aab}, where the associated
$L_\infty$ algebra structure was 
derived from a Borcherds superalgebra $\scr B(\fgrplus)$.
In the present paper, we extend the analysis and include the cases
where ancillary transformations are present 
by using a THA $S(\fgrplus)$ rather than $\scr
B(\fgrplus)$. 
Accordingly, also the study of the THA's, restricted 
to finite-dimensional $\fgrplus$ in \cite{Carbone:2018xqq}, needs to be extended.
This extended study is carried out in the accompanying paper \cite{CederwallPalmkvistTHAI}.

Throughout the paper, all considerations concern what in the context
of \eg\ double or extended geometry, or the geometrisation of Ehlers
symmetry, would be the ``internal'' sector. Most situations we
describe do not have an interpretation in terms of string theory or
M-theory duality, although geometrisation of any duality arising on
compactification to $3$ dimensions or higher is covered. Dualities in
$3$ dimensions generically correspond to extended geometry with the
adjoint as coordinate representation. This is a subclass of the models
described in this paper, but not included in the analysis of ref.
\cite{Cederwall:2018aab}.

The paper is organised as follows. Section \ref{ReviewSection}
contains a brief review of earlier work on the gauge structure and
dynamics of extended geometry in cases where $(\lambda,\theta)=1$.
Section \ref{THASection} describes the THA
$S(\fgrplus)$ with focus on a particularly useful double
grading. For more detail, we refer to the companion paper
\cite{CederwallPalmkvistTHAI}.
We then perform a detailed investigation of the ancillary
transformations, as obtained from the THA,
in Section \ref{AncillarySection}. The dynamics is formulated by
means of a pseudo-action, using structure constants of the THA,
in Section \ref{DynamicsSection}, and the
$L_\infty$ gauge structure is outlined in Section \ref{LinftySection}.
We conclude with a discussion in Section \ref{DiscussionSection}, with
focus on the continuation to infinite-dimensional structure algebras
$\fgr$.

The accompanying paper \cite{CederwallPalmkvistTHAI} deals with the
construction of the THA's from generators and
relations, and other purely algebraic aspects. In order for both
papers to be reasonably self-contained, their contents have a certain overlap. 

\section{Review of extended geometry\label{ReviewSection}}

The input for extended geometry \cite{Cederwall:2017fjm}
is a structure group $G$ (with Lie
algebra $\fgr$) together with a lowest weight ``coordinate
representation'' (\ie, the representation of a generalised tangent vector)
$R(-\lambda)$. The generalised tangent space is the lowest weight
module $R(-\lambda)$, where $\lambda$ is an integral dominant weight.
The structure algebra can in principle be any Kac--Moody algebra, but
we will, for simplicity, take $\fgr$ to be simply laced (or at least that $\lambda_i=0$ when $\alpha_i$ is a short root) and normalise the simple roots to have
length squared $2$.

\begin{figure}
  \centerline{\includegraphics{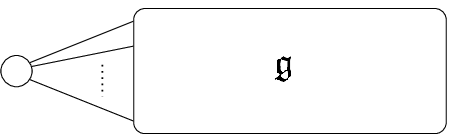}}
  \caption{\it The Dynkin diagram for $\fg^+$.}
  \label{DynkingplusFig}
\end{figure}

In what follows, we often use an extension of $\fg$ to a Lie algebra
$\fg^+$, which is obtained by adjoining one node, corresponding to a simple
root with length squared $2$, to the Dynkin diagram of $\fg$, with lines
corresponding to the coefficients of $\lambda$ expressed in the basis of
fundamental weights, see Figure \ref{DynkingplusFig}.

\subsection{Generalised diffeomorphisms and ancillary transformations} 

Generalised diffeomorphisms with infinitesimal parameter $\xi\in
R(-\lambda)$, acting on a vector $V\in
R(-\lambda)$, are given in terms of the
generalised Lie derivative,
\begin{align}
\LL_\xi V^M=\xi^N\*_NV^M+Z_{PQ}{}^{MN}\*_N\xi^PV^Q\;.
\end{align}
The second term is a local $\fgr\oplus\mathbb{R}$ transformation of $V$.
The invariant tensor $Z$ has the universal expression
\begin{align}
\sigma Z=-\eta^{\alpha\beta}t_\alpha\otimes t_\beta+(\lambda,\lambda)-1\;,
\end{align}
where $\sigma$ is the permutation operator, $t_{\alpha M}{}^N$
representation matrices, and $\eta$ the inverse Killing metric.
Here, and in the following, we use an index-free notation, where $Z$
is seen as an operator 
$Z:\,R(\lambda)\otimes R(\lambda)\rightarrow R(\lambda)\otimes R(\lambda)$.
The Killing metric and its inverse will often be used to implicitly
raise and lower adjoint indices.  
Note that the generalised Lie derivative fulfils a Leibniz rule.

Calculating the commutator of two generalised diffeomorphisms, and
using the section constraint gives the result \cite{Cederwall:2017fjm}
\begin{align}\label{GenDiffCommutator}
[\LL_\xi,\LL_\eta]V=(\LL_{\frac12(\LL_\xi\eta-\LL_\eta\xi)}+\Sigma_{\xi,\eta})V\;.
\end{align}
The second
term on the right hand side is a so-called ancillary transformation,
which is a restricted local $\fgr$ transformation.

The calculation leading to \eqref{GenDiffCommutator} uses
the section constraint. This constraint ensures that any two
derivatives on any field or parameter lies in a linear subspace (a
section) of the
minimal orbit of $R(\lambda)$ under $G$. The stability group of a
section is a parabolic subgroup of $G$ containing $GL(d)$,
the local structure group of gravity.
The section constraint reads
\begin{align}
Y\*\otimes\*=0\;,
\end{align}
where $Y=Z+1$, \ie,
\begin{align}
\sigma Y=-\eta^{\alpha\beta}t_\alpha\otimes t_\beta+(\lambda,\lambda)-1+\sigma\;.
\end{align}
The relation of the section constraint to minimal orbits was
elaborated in refs. \cite{Bossard:2017wxl,Cederwall:2017fjm}.
On a state\footnote{We use the notation $\vee$ for the symmetrised
  and $\wedge$ for the anti-symmetrised product.}
$\phi\in R(2\lambda)\subset \vee^2R(\lambda)$, one
has
\begin{align}
  \eta^{\alpha\beta}t_\alpha\otimes
  t_\beta\,\phi=(\lambda,\lambda)\phi\;,
\end{align}
and on $\psi\in R(2\lambda-\alpha_i)\subset \wedge^2R(\lambda)$, where
$(\lambda,\alpha_i)=1$,
\begin{align}
  \eta^{\alpha\beta}t_\alpha\otimes
  t_\beta\,\psi=((\lambda,\lambda)-2)\psi\;.
\end{align}
These relations are straightforward to derive using the eigenvalues of the
quadratic Casimir operator in the respective representations.
The representations $R(2\lambda)$ (symmetric) and
$\bigoplus_{i:\lambda_i=1}R(2\lambda-\alpha_i)$ (antisymmetric) are the
representations in $\otimes^2R(\lambda)$ which are 
annihilated by $Y$.

The ancillary transformation $\Sigma_{\xi,\eta}$ has the generic form
\cite{Cederwall:2017fjm}
\begin{align}\label{SigmaEQ}
  \Sigma_{\xi,\eta}V^M
  =\frac12S^\alpha{}_{PQ}{}^{RS}\xi^P\*_R\*_S\eta^Q
  t_{\alpha N}{}^MV^N-(\xi\leftrightarrow\eta)\;,
\end{align}
the tensor $S$ being given by the simple expression\footnote{In
  ref. \cite{Cederwall:2017fjm}, 
  the projection was not included in the
  definition of the tensor $S$. In addition, there is an apparent sign
difference to the expression given there; this is due to the change of
conventions for the $t_\alpha$'s.}
\begin{align}
  \label{STensorEq}
  S^\alpha{}_{MN}{}^{PQ}=
  \left(f^{\alpha\beta\gamma}t_\beta\otimes t_\gamma+t^\alpha\otimes\id
  -\id\otimes t^\alpha\right){}_{MN}{}^{\langle PQ\rangle}\;.
\end{align}  
The notation $\langle PQ\rangle$ means projection on $R(2\lambda)$ or
$R(-2\lambda)$.
Later, we will also denote the projection on
the antisymmetric $\bigoplus_{i:\lambda_i=1}R(2\lambda-\alpha_i)$ or
its conjugate by a
curly bracket $\{MN\}$.
The tensor $S$ is antisymmetric in its lower indices, and obeys
$S^\alpha{}_{\{MN\}}{}^{PQ}=0$.

The tensor $S$ vanishes when $\fg$ is a finite-dimensional algebra and
$\lambda$ is a weight with $(\lambda,\theta)=1$, \ie, a fundamental
weight dual to a simple root with Coxeter label $1$.
In ref. \cite{Cederwall:2017fjm} we gave the following list of cases
where ancillary transformations are absent: 
\begin{itemize}
\item
{$\fgr=A_r$, $\lambda=\Lambda_p$, $p=1,\ldots,r$
($p$-form representations);}
\item 
{$\fgr=B_r$, $\lambda=\Lambda_1$ (the vector representation);}
\item 
{$\fgr=C_r$, $\lambda=\Lambda_r$ (the 
symplectic-traceless $r$-form representation);}
\item 
{$\fgr=D_r$,
$\lambda=\Lambda_1,\Lambda_{r-1},\Lambda_r$ (the vector and spinor
representations);}
\item 
{$\fgr=E_6$, $\lambda=\Lambda_1,\Lambda_5$ (the
fundamental representations);}
\item 
{$\fgr=E_7$, $\lambda=\Lambda_1$ (the fundamental
representation).}
\end{itemize}
In many of these cases, $\fgrplus$ is finite-dimensional, but the list
also contains cases where $\fgrplus$ is infinite-dimensional: 
\begin{itemize}
\item 
$\fgr=A_7$, $\lambda=\Lambda_4$;
\item 
$\fgr=A_r$, $r\geq 8$, $\lambda=\Lambda_p$, $p=3,\ldots,r-2$;
\item 
$\fgr=C_r$, $r\geq 4$, $\lambda=\Lambda_r$;
\item 
$\fgr=D_r$, $r\geq 8$,
$\lambda=\Lambda_{r-1},\Lambda_r$.
\end{itemize}

Assuming that $\fg$ is simply laced, the presence of ancillary
transformations implies that $\fg^+$ is infinite-dimensional.
We will not consider the cases where $\fg$ is not simply laced further
in the present paper,
but the construction should work as long as $\lambda$ is orthogonal to
all short roots.

\subsection{Dynamics}

In ref. \cite{Cederwall:2017fjm}, a (pseudo-)action\footnote{We are
  reluctant to use the term action as long as the section constraint
  needs to be imposed by hand.} was given,
encoding the dynamics of any extended geometry in the absence of
ancillary transformations, and in the special cases where the
coordinate representation is the adjoint.

The generalised metric $G_{MN}$ parametrises the coset
$G/K\times{\mathbb R}$. It is convenient, in order
to write an action without reference to any ``external'' dimensions,
to let $G_{MN}$ transform as
a tensor density with weight $-2w=1-2(\lambda,\lambda)$. The
generalised metric induces a
preferred involution on $\fgr$ through
$t_{\alpha M}{}^N\mapsto-t^\star_{\alpha M}{}^N=-(G^{-1}t_\alpha
G)^N{}_M$, \ie, $t^\star=Gt^tG^{-1}$.

Define
\begin{align}
  (\*_MGG^{-1})_N{}^P=\Pi_{\alpha M}t^\alpha{}_N{}^P+\Pi_M\delta_N^P\;
\end{align}
(this decomposition follows from $G$ being a group element in
$G\times\RR$).
Checking the transformation of a Lagrangian up to total derivatives, only
the inhomogeneous transformations $\Delta_\xi\equiv\delta_\xi-\LL_\xi$
are needed. They are
\begin{align}
\Delta_\xi\Pi_M&=-2w\*_M\*_N\xi^N\;,\nn\\
\Delta_\xi\Pi_{\alpha M}
&=(t_\alpha+t^\star_\alpha)_P{}^N\*_M\*_N\xi^P\;.
\end{align}
It was then shown that the combination of terms
\begin{align}
\LL_0=\fr2A-B-2C
-{(\lambda,\lambda)\over(\lambda,\lambda)-\fr2}D\;,
\end{align}
with
\begin{align}
A&=G^{MN}\eta^{\alpha\beta}\Pi_{\alpha M}\Pi_{\beta N}\;,\nn\\
B&=G^{PQ}t^\alpha{}_P{}^Mt^\beta{}_Q{}^N\Pi_{\alpha N}\Pi_{\beta
  M}\;,\nn\\
C&=( G^{-1}t^\alpha)^{MN}\Pi_M\Pi_{\alpha N}\;,\nn\\
D&=G^{MN}\Pi_{ M}\Pi_{ N}\;,
\end{align}
has the inhomogeneous transformation, up to a total derivative,
\begin{align}\label{RemainderTerm}
\Delta_\xi\LL_0=-2S^\alpha{}_{PQ}{}^{MN}G^{PS}\Pi_{\alpha
    S}\*_M\*_N\xi^Q\;,
\end{align}
The calculation relies on the section condition, both its symmetric
part on the two derivatives on $\xi$ and its general form on $\Pi$ and
one of the derivatives,
in order to cancel terms produced by the $\sigma$ term in the $Y$
tensor, and thereby relating terms with adjoint and scalar parts of $\Pi$.
The weight of each derivative is $-(\lambda,\lambda)+1$ and that of
an inverse metric $2(\lambda,\lambda)-1$, so each term in the
Lagrangian has weight $1$. This is the correct weight for a density $\omega$ so
that being a divergence $\omega=\*_Mv^M$ of a vector density $v$ is
a covariant property \cite{Cederwall:2017fjm}.
The Lagrangian $\LL_0$ thus describes the dynamics in the absence of
ancillary transformations.

\subsection{Gauge structure, $L_\infty$ and the Borcherds superalgebra
$\sB$\label{LInftySubSection}}

In ref. \cite{Palmkvist:2015dea,Cederwall:2018aab}, the generalised tangent space
was identified with a subspace of a Borcherds superalgebra, here denoted $\scr B(\fg^+)$, or simply just $\BB$. 
This superalgebra is obtained by a double extension
of the structure algebra $\fgr$, first to $\fgrplus$
by a node corresponding to the
weight $\lambda$, attached to node $i$ in the Dynkin diagram of $\fgr$ with a number of lines equalling the Dynkin labels $\lambda_i$ of $\lambda$, and
then by a ``grey''
node attached with a single line to the first extending node. The simple root corresponding to the grey node is an odd null root.
The resulting Dynkin diagram is shown as the right diagram in Figure
\ref{DynkinFig}. The left diagram in the same Figure is equivalent.
The Borcherds superalgebra $\BB$ has a consistent ($\mathbb{Z}\times\mathbb{Z}$)-grading, corresponding to the two
leftmost nodes in the left diagram. We 
denote the corresponding grades by $p$ and $q$ (where $q$ corresponds to the leftmost node) and refer to them as {\it level} and {\it height},
respectively.

Our notation for the basis elements in the local part of the algebra
with respect to the level $p$ is given in Table
\ref{GeneralBTableBasis}.
At height $q=0$, we have the subalgebra
$\BB(\fgr)\oplus\RR$, and at $p=q$ the subalgebra
$\fgrplus\oplus\RR$. Elements at a given ($\mathbb{Z}\times\mathbb{Z}$)-grade form $\fgr$-modules. The modules at even and odd $p+q$ belong to the even and odd parts of the Lie superalgebra, respectively. We refer to refs. \cite{Cederwall:2018aab,CederwallPalmkvistTHAI} for
more detail.

The subspace where the vector fields live is the one at $(p,q)=(1,0)$, with basis $E_M$, but in order to describe
the gauge structure it is convenient to define a generalised Lie derivative for any pair of elements $A$ and $B$ at $p\geq0$ and $q=0$, which are also allowed to have odd (fermionic) components.
The generalised Lie derivative is then constructed
\cite{Palmkvist:2015dea,Cederwall:2015oua,Cederwall:2017fjm,Cederwall:2018aab}
as
\begin{align}
\LL_AB=\delta_{p_A,1}\left([[A,F^{\fl M}],\*_MB]
     +(-1)^{|B|}[[\*_MA^\sh,F^{\fl M}],B]\right)\;,
\end{align}
where $|B|$ is the total statistics of $B$ and $p_A$ is the level of $A$. The bracket $[\cdot,\cdot]$ is that of the Borcherds superalgebra
$\BB(\fgrplus)$ (we use this notation also for the symmetric bracket
between two totally fermionic elements).

The ghosts $C_p$ reside at height $q=0$, and the
ancillary ghosts $K_p$ at $q=1$, see Table
\ref{DerivativeActionTable}. 
They can be combined into $C=\sum_{p=1}^\infty C_p$ and $K=\sum_{p=p_0}^\infty K_p$.
Both the ghosts and ancillary ghosts are totally bosonic, $|C|=|K|=0$.
The ghost number is $p+q$.
Note that $p_0\geq1$, which means that the Borcherds superalgebra can
not accommodate ancillary transformations, corresponding to ancillary
ghosts with ghost number $1$, which would reside at
$(p,q)=(0,1)$ and transform as the adjoint. This will be remedied by
the introduction of the tensor 
hierarchy algebras.

\begin{table}
\begin{align*}
  \xymatrix{
    &&&&K_{p_0}\ar[d]_\fl\ar@{<-}[r]_d
    &K_{p_0+1}\ar[d]_\fl\ar@{<-}[r]_d
    &K_{p_0+2}\ar[d]_\fl\ar@{<-}[r]_d&\cdots\\
    &C_1\ar@{<-}[r]_d
    &\cdots\ar@{<-}[r]_d
    &C_{p_0-1}\ar@{<-}[r]_d
    &C_{p_0}\ar@{<-}[r]_d
    &C_{p_0+1}\ar@{<-}[r]_d
    &C_{p_0+2}\ar@{<-}[r]_d&\cdots\\
}
\end{align*}
\caption{\it The typical structure of the action of the $1$-bracket
  between the ghost modules,
  with ancillary ghosts appearing from level $p_0\geq1$.}
\label{DerivativeActionTable}
\end{table}

We denote $C=c+C'$, where $c=C_1$. In ref. \cite{Cederwall:2018aab} it was shown that the ghosts satisfy an $L_\infty$ algebra \cite{Lada:1992wc,Zwiebach:1992ie,Hohm:2017pnh,Roytenberg:1998vn}. The full 
list of non-vanishing brackets is:
\begin{align}\label{FullListBorcherds}
\dlb C'\drb&=dC'\;,\nn\\
\dlb K\drb&=dK+K^\fl\;,\nn\\
\dlb C^n\drb&=k_n\Bigl((\ad C)^{n-2}(\LL_CC+X_CC)
+\sum\limits_{i=0}^{n-3}(\ad C)^iR_C(\ad C)^{n-i-3}\LL_CC\Bigr)\;\nn\\
\dlb C^{n-1},K\drb&=\frac{k_n}n\Bigl(
  (\ad C)^{n-2}\LL_CK+
  \sum\limits_{i=0}^{n-3}(\ad C)^i\ad K(\ad C)^{n-i-3}\LL_CC
  \Bigr)\;,
\end{align}
where the coefficients $k_n$ have the universal model-independent expression
in terms of Bernoulli numbers
\begin{align}
  k_{n+1}=\frac{2^nB^+_n}{n!}\;,\quad n\geq1\;.
\end{align}
We here use the convention that the brackets are graded symmetric, all brackets carry ghost number $-1$, and ``$C^n$'' represents ``$C,C,\ldots,C$''.
All non-vanishing 
brackets except the $1$-bracket contain at least one level $1$ ghost $c$.
No brackets contain more than one ancillary ghost.

The meaning of the symbols in eq. \eqref{FullListBorcherds} will
be explained in detail in Section \ref{LinftySection}.
Here, it suffices to mention that $X$ and $R$ represent ancillary
contributions, and that the nilpotent
``derivatives'' $d$ and $\flat$ act to the left and down,
respectively (see Table \ref{DerivativeActionTable}),
in the double grading of Table \ref{GeneralBTableBasis}.

\section{The tensor hierarchy algebra $S$\label{THASection}}

When $\fgr=A_{n-1}$, \ie, when extended geometry is geometry with
structure group $GL(n)$, the
tensor hierarchy algebras are $W(A_{n-1})=W(n)$, and
$S(A_{n-1})=S(n)$, the
finite-dimensional non-contragredient superalgebras of Cartan type in Kac's
classification \cite{Kac77B}. The definition of these superalgebras 
in terms of generators and relations derived from a Dynkin
diagram---in spite of them being non-contragredient---was given in
ref. \cite{Carbone:2018xqq}. This construction is extended in the
accompanying paper \cite{CederwallPalmkvistTHAI} to $W(\fgrplus)$ and
$S(\fgrplus)$ for finite-dimensional $\fg$.

\begin{figure}
  \centerline{\includegraphics{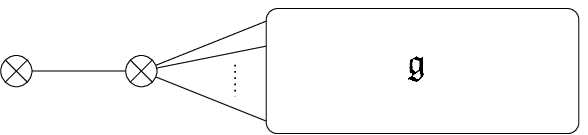}\hskip5mm\includegraphics{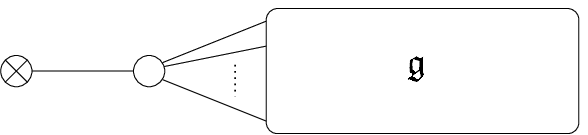}}
  \caption{\it Two equivalent Dynkin diagrams for $\BB(\fgrplus)$, $W(\fg^+)$
    and $S(\fgrplus)$.}
  \label{DynkinFig}
\end{figure}

Here we will focus on $S=S(\fgrplus)$, where $\fgrplus$ is obtained,
as explained above, by
adding one node to a finite-dimensional Lie algebra $\fgr$, which
is the structure algebra of the extended geometry under consideration.
The superalgebra $S$ turns out to be appropriate for the
description of extended geometry; the extra elements in $W=W(\fgrplus)$
are not required.
Both $W$ and $S$ can be described by the same Dynkin diagram as $\sB$,
but with different assignment of generators and relations
\cite{CederwallPalmkvistTHAI}.
This doubly extended diagram was described in the preceding Section and is given schematically in Figure
\ref{DynkinFig}. 

For the construction of gauge transformations etc. (and ultimately the
$L_\infty$ algebra), we use the same double grading as for $\BB$, splitting the
superalgebra into finite-dimensional representations of $\fgr$,
labelled by two integers $(p,q)$. The subalgebra $W(\fgr)\subset
S(\fgrplus)$ is on the line $q=0$, and the Lie subalgebra
$\fgrplus\subset S(\fgrplus)$ is on the
diagonal $p=q$ (generically together with other elements).

\begin{table}
  \begin{align*}
  \xymatrix@=.4cm{
    \ar@{-}[]+<2.2em,1em>;[ddd]+<2.2em,-1em>
    \ar@{-}[]+<-0.8cm,-1em>;[rrr]+<1.4cm,-1em>
    &\ar@{-}[]+<2.3em,1em>;[ddd]+<2.3em,-1em> p=-1
    & \ar@{-}[]+<2.9em,1em>;[ddd]+<2.9em,-1em> p=0 &p=1\\
q=1&&f_0& E_M^\sh
       \\ 
q=0&F^M
&k\quad{T_\alpha}\quad\tk
           & E_M\\ 
q=-1 & F^{\fl M}& e_0 
  }
\end{align*}
  \caption{\it Basis elements for $\sB$ at $p=-1,0,1$.}
\label{GeneralBTableBasis}
\end{table}

\begin{table}
  \begin{align*}
  \xymatrix@=.4cm{
    \ar@{-}[]+<2.2em,1em>;[dddd]+<2.2em,-1em>
    \ar@{-}[]+<-0.8cm,-1em>;[rrr]+<1.4cm,-1em>
    &\ar@{-}[]+<3.3em,1em>;[dddd]+<3.3em,-1em> p=-1
    & \ar@{-}[]+<2.4em,1em>;[dddd]+<2.4em,-1em> p=0 &p=1\\
q=2&&&L^\sh_{\alpha M}\\
q=1&\Phi_{\alpha}^{\sh M}&T_\alpha^\sh& E_M^\sh\quad L_{\alpha M}
       \\ 
q=0&\Phi_{\alpha}{}^M\quad H^M
&{T_\alpha}\quad\tk
           & E_M\\ 
q=-1 & H^{\fl M}& {e_0} 
  }
\end{align*}
  \caption{\it Basis elements for $S$ at $p=-1,0,1$.}
\label{GeneralSTableBasis}
\end{table}

\begin{table}
  \begin{align*}
  \xymatrix@=.4cm{
    \ar@{-}[]+<2.2em,1em>;[dddd]+<2.2em,-1em>
    \ar@{-}[]+<-0.8cm,-1em>;[rrr]+<1.4cm,-1em>
    &\ar@{-}[]+<4.4em,1em>;[dddd]+<4.4em,-1em> p=-1
    & \ar@{-}[]+<3em,1em>;[dddd]+<3em,-1em> p=0 &p=1\\
q=2&&&L^\sh_{\alpha M}\\
q=1&\Phi_{\alpha}^{\sh M}\quad G^{\sh M}&f_0\quad T_\alpha^\sh& E_M^\sh\quad L_{\alpha M}
       \\ 
q=0&F^M\quad\Phi_{\alpha}{}^M\quad G^M
&k\quad T_\alpha\quad\tk
           & E_M\\ 
q=-1 & F^{\fl M}& {e_0} 
  }
\end{align*}
  \caption{\it Basis elements for $W$ at $p=-1,0,1$.}
\label{GeneralWTableBasis}
\end{table}

A way of deriving the content of $S$ is to note
that the generators $E_M$, which form a basis for $R_1=R(-\lambda)$,
have a covariant Serre relation in
$R(-2\lambda)$, so that the anticommutator $[E_M,E_N]$ lies in
$R_2=\vee^2R_1\ominus R(-2\lambda)$.
Any element at $(p,q)=(-1,0)$ must respect the ideal generated by the
Serre relations.
This allows for the introduction of generators $\Phi_\alpha{}^M$ 
with anticommutators
\begin{align}
[E_N,\Phi_\alpha{}^M]=\varphi^\beta{}_{N,\alpha}{}^MT_\beta\;,
\end{align}
where $\varphi$ is a linear combination of projection operators on the
irreducible modules appearing in $\varphi$. 
They respect the ideal in $R(-2\lambda)$ if
\begin{align}
  \label{EmbeddingTensorCondition}
  t_{\beta\langle M}{}^P\varphi^\beta{}_{N\rangle,\alpha}{}^Q=
  (t_\beta\otimes\varphi^\beta{}_\alpha)_{\langle MN\rangle}{}^{PQ}=0\;,
\end{align}
or equivalently,
$(\phi^\alpha{}_\beta\otimes t^\beta)_{MN}{}^{\langle PQ\rangle}=0$.
where $\langle MN\rangle$ is projection on $R(\pm2\lambda)$.
Eq. \eqref{EmbeddingTensorCondition} is the condition for the
representation of the embedding tensor, or the ``big torsion
representation''\footnote{Although we have not performed a complete
  analysis, we have noted that in cases when $\lambda$ is attached to
  a short root, there is typically no solution to this algebraic
  condition.}
The content of $\Phi$ in terms of irreducible representations can be determined by decomposing $\fg$ into levels $\ell$ with respect to $\lambda$
(so that the level of a root $\alpha$ is $(\lambda,\alpha)$) \cite{CederwallPalmkvistTHAI}. If we then let $H_0$ be the set of highest roots at level $0$ and $L_\ell$ the
set of lowest roots at level $\ell$, then the content of $\Phi$ is
\begin{align}
  (N-1)R(\lambda)\oplus\bigoplus\limits_{\gamma\in H_0}R(\lambda+\gamma)
  \oplus\bigoplus\limits_{\ell=2}^{(\lambda,\theta)}
  \bigoplus\limits_{\beta\in L_\ell}R(\lambda-\beta)\;,
  \label{PhiModules}
\end{align}
where $N$ is the number of
non-vanishing Dynkin labels of $\lambda$.

The difference at $p=1$ between Tables \ref{GeneralBTableBasis} and
\ref{GeneralSTableBasis}
illustrates the shortcoming of $\sB(\fgrplus)$, namely that
it does not contain a module $\tR_1$, and thus is unable to encode ancillary
transformations (ancillary ghosts with ghost number $1$).
The THA's, on  the other hand, contain $\tR_1$
precisely when ancillary transformations occur --- the corresponding
generators are $L_{\alpha M}$ and $L^\sh_{\alpha M}$ in Table
\ref{GeneralSTableBasis}.
There are of course further differences. At $p=0$, $q=1$, the THA
has an adjoint element, which is where the ancillary ghost
will reside. At $p=-1$, there is also room for generalised torsion.
We comment on
the inclusion of dynamical variables in the discussion in Section
\ref{DiscussionSection}.

In a single grading with respect to the
  leftmost node in the second diagram of Figure~\ref{DynkinFig}, \ie, where
  the grading where each grade $n=p-q$ forms a $\fgrplus$-module, the positive levels
of $W$
agree
with those of the corresponding Borcherds superalgebra $\BB$. Also the positive levels of $S$ essentially agree with those of
$\BB$, but 
in
  some cases, there
  may be an issue of further ideals arising at positive levels in the
  definition of the 
  algebra $S$ \cite{CederwallPalmkvistTHAI}. Up to such ideals (further discussed in ref. \cite{CederwallPalmkvistTHAI}),
both $S$ and $\sB$ are subalgebras of
$W$, with the grading shown in Table
\ref{GeneralWTableBasis}. The elements in the subalgebras denoted by
the same symbols as in $W$ are identified. In addition, the
elements $F^\fl$ and $H^\fl$ are the same, although $F$ and $H$ are
not; $H$ is the linear combination\footnote{The special case
  $(\lambda,\lambda)=1$ requires a different treatment. In that case
$W(\fgr)\not\subset S(\fgrplus)$. This happens for $\fgr=D_r$,
  $\lambda=\Lambda_1$, \ie, for double geometry.
  Since no ancillary ghosts appear, we have not
  investigated this case further.}
$H^M=F^M+\frac1{(\lambda,\lambda)}G^M$.
The criterion determining this
combination (apart from $H^\fl=F^\fl$) is that the bracket $[E_M,H^N]$
may contain $\tk$ but not $k$.
The lowering operation is defined as $A^\fl=-[A,e_0]$ on any
element. The raising operator is defined so that
$A^\sh=0$ for an upper element in a pair, and
$A^{\fl\sh}+A^{\sh\fl}=A$. Although we use the same notation
for raising and lowering as for the Borcherds superalgebra in
ref. \cite{Cederwall:2017fjm}, these are different operations (which
is illustrated by $H$ and $H^\fl$ above). The
ones in the Borcherds superalgebra are inherited from $W$,
where raising is not defined at $p=0$. On the other hand, in $S$, raising and lowering are
well defined operations at level $p=0$,
but it is not completely clear how to define the raising operator in general.

We refer to ref. \cite{CederwallPalmkvistTHAI} for the mathematical
details concerning the definition and
construction of $S$. For the class
of algebras considered here, the representation content of $\tR_1$ can
be explicitly determined,
\begin{align}
  \label{tRContent}
  \tR_1=\bigoplus\limits_{\ell=2}^{(\lambda,\theta)}
  \bigoplus\limits_{\beta\in L_\ell}R(-(\lambda-\beta))\;.
\end{align}
Note that the earlier criterion
for the absence of ancillary transformations \cite{Cederwall:2017fjm},
namely $(\lambda,\theta)=1$, implies that the grading is a 3-grading
with $\ell=-1,0,1$,
and the sum in \eqref{tRContent} becomes empty.
The simplest class of examples with non-vanishing $\tR_1$ is when
$\lambda=\theta$, so that $R_1$ is the adjoint of $\fg$. Then
$(\lambda,\theta)=2$, and the sum contains a single irreducible
representation, the singlet.

It is a peculiar property of the THA's, that when
they are the super-extension of an infinite-dimensional Kac--Moody
algebra (here, $\fgrplus$), already the level $n=0$ (in the double
grading, $p-q=0$) contains elements beyond the adjoint of $\fgrplus$ \cite{Bossard:2017wxl}.
This property follows \cite{CederwallPalmkvistTHAI} from the natural
generalisation of the identities for the generators associated to the
Dynkin diagram given in ref. 
\cite{Carbone:2018xqq}. It also provides exactly the structure needed
for the description of the ancillary transformations, which will be
explained in the following section. See
ref. \cite{CederwallPalmkvistTHAI} for more detail.

Some non-trivial brackets in $S$ include
\begin{align}
  [\tk,E_M]&=(1-(\lambda,\lambda))E_M\;,\nn\\
  [\tk,e_0]&=-e_0\;,\nn\\
  [E_M,H^N]&=-\left(1-\frac1{(\lambda,\lambda)}\right)t^\alpha{}_M{}^NT_\alpha
  +\delta_M^N\tk\;,\nn\\
  [E_M,H^{\fl N}]&=\delta_M^Ne_0\;,\nn\\
  [E^\sh_M,H^{\fl N}]&=-t^\alpha{}_M{}^NT_\alpha
  +\delta_M^N\tk\;,\nn\\
  [E^\sh_M,H^N]&=-\frac1{(\lambda,\lambda)}t^\alpha{}_M{}^NT^\sh_\alpha\;,
  \label{TshEBracket}\nn\\
  [T^\sh_\alpha,H^{\fl M}]&=t_{\alpha N}{}^MH^N+\Phi_\alpha{}^M\;,\nn\\
  [T^\sh_\alpha,E_M]&=t_{\alpha M}{}^NE^\sh_N+L_{\alpha M}\;.
\end{align}
The last two of these can be taken as definitions of $\Phi$ and
$L$, containing the modules \eqref{PhiModules} and \eqref{tRContent},
respectively.  
We also postulate
\begin{align}
  [E_N,\Phi_\alpha{}^M]&=\varphi^\beta{}_{N,\alpha}{}^MT_\beta\;,\nn\\
  [L_{\alpha M},H^{\fl N}]&=\ell_{\alpha M}{}^{\beta N}T_\beta\;.
  \label{EllTensor}
\end{align}
These relations serve to define the $\fgr$-invariant tensors
$\varphi$ and $\ell$, which by construction are some linear combinations
of projectors on the irreducible 
modules in $\Phi$ and $L$.

Using the commutators, one can also check explicitly that $\Phi$ respects the
ideal $\bigoplus_{i:\lambda_i=1}R(-(2\lambda-\alpha_i))$ in
$[E^\sh_M,E^\sh_N]$. The condition becomes
\begin{align}
L^\sh_{\beta\{N}\varphi^\beta{}^{\mathstrut}_{P\},\alpha}{}^M=0\;.
\end{align}
This is automatically satisfied, since the highest representations in
$\varphi$ and $\ell$ are $R(\lambda+\gamma_0)$ and $R(\lambda-\beta_2)$,
where $\gamma_0$ is a highest root at level $0$. The tensor product
can not contain $R(2\lambda-\alpha_i)$.

Consider the Jacobi identity between $T^\sh_\alpha$, $E_M$
and $H^{\fl N}$. This turns out to be the only non-trivial Jacobi
identity within  the local superalgebra at $p=-1,0,1$, in the sense
that all others can be obtained from it by raising and lowering
operations.
A short calculation leads to the necessary and sufficient condition
for this Jacobi identity to be fulfilled:
\begin{align}  
  \varphi^\beta{}_{M,\alpha}{}^N-\ell_{\alpha M}{}^{\beta N}
  =\delta_\alpha^\beta\delta_M^N
  -f_\alpha{}^{\beta\gamma}t_{\gamma M}{}^N
  -\frac1{(\lambda,\lambda)}(t^\beta t_\alpha)_M{}^N
  &\equiv Q_{\alpha M}{}^{\beta N}\;,
  \label{PhiLRelation}
\end{align}
\ie,
\begin{align}
  \varphi^ \beta{}_\alpha-\ell_\alpha{}^\beta=\delta_\alpha^\beta
  -f_\alpha{}^{\beta\gamma}t_\gamma-\frac1{(\lambda,\lambda)}t^\beta
  t_\alpha
  \equiv Q_\alpha{}^\beta\;.
  \label{PhiLRelation2}
\end{align}
If we now make use of the algebraic condition
\eqref{EmbeddingTensorCondition} on $\varphi$, the part of
this relation only  involving $\ell$ becomes, using the section constraint,
\begin{align}
  \ell_{\beta M}{}^{\alpha\langle P}t^\beta{}_N{}^{Q\rangle}
  &=f^\alpha{}_{\beta\gamma}t^\beta{}_M{}^{\langle P}t^\gamma{}_N{}^{Q\rangle}
  +t^\alpha{}_M{}^{\langle P}\delta_N^{Q\rangle}
  -\delta_M^{\langle P}t^\alpha{}_N{}^{Q\rangle}\nn\\
  &=(f^\alpha{}_{\beta\gamma}t^\beta\otimes t^\gamma+t^\alpha\otimes\id
  -\id\otimes t^\alpha)_{MN}{}^{\langle PQ\rangle}
  \;.
  \label{EllTIsS}
\end{align}
We recognise the right hand side as the $S$ tensor
of eq. \eqref{STensorEq}, and thus
\begin{align}
  (\ell_\beta{}^\alpha\otimes t^\beta)_{MN}{}^{\langle PQ\rangle}
  =S^\alpha{}_{MN}{}^{PQ}\;.
  \label{EllTIsS2}
\end{align}
The tensor $S$ is antisymmetric in its lower
indices. In addition, it satisfies
$S_{\{MN\}}{}^{PQ}=0$, thanks to the identity
\begin{align}
  S^\alpha{}_{MN}{}^{PQ}
=\Bigl(\frac{1-\sigma}2Y(\id\otimes t^\alpha)\Bigr)_{MN}{}^{\langle
  PQ\rangle}\;.
\end{align}

We would ideally like to show that there always is a solution of this form to
eq. \eqref{PhiLRelation2}. This follows from the existence of the
THA as defined in
ref. \cite{CederwallPalmkvistTHAI}, but seems surprisingly difficult
to prove in a more direct manner, only using representation theory for $\fg$.
The difficulty with analysing this equation lies in the translation
between the projectors on irreducible representations in
$\adj\otimes R(\lambda)$ of the types $P_{\alpha M}{}^{\beta N}$ and
$P^\beta{}_{M,\alpha}{}^N$‚ used to characterise $\ell$ and $\varphi$,
respectively, which are not known explicitly (in any useful form)
in the general case.

In ref. \cite{CederwallPalmkvistTHAI}, we discuss the remarkable
identity \eqref{PhiLRelation2} in more detail, and show that the
matrix $Q$ on the right 
hand side in a certain sense has corank $2$, which makes the solution
in terms of $\varphi$ and $\ell$ possible for any integrable highest
weight representation of a finite-dimensional simply laced $\fg$.
Some highly non-trivial examples of this relation
are also given in ref. \cite{CederwallPalmkvistTHAI}.

\section{Ancillary transformations from $S$\label{AncillarySection}}

We can  perform the calculation of the ancillary term in the
commutator of two generalised diffeomorphisms with the expressions for the
derived brackets directly in terms of the superalgebra brackets. The expression
for the generalised Lie derivative is identical to the one using the
Borcherds superalgebra \cite{Cederwall:2018aab} (but there the
ancillary transformations could not be derived in terms of the
superalgebra brackets).

Let
\begin{align}
\LL_\xi V=[[\xi,H^{\fl M}],\*_MV^\sh]-[[\*_M\xi^\sh,H^{\fl M}],V]\;,
\end{align}
where $\xi$ has bosonic components, and where
$\xi=\xi^ME_M$, $V=V^ME_M$.
Consider the $\xi\*^2\eta$ terms in
$[\LL_\xi,\LL_\eta]V-\LL_{\frac12(\LL_\xi\eta-\LL_\eta\xi)}V$. They
become
\begin{align}
  &-[[\xi,H^{\fl M}],[[\*_M\*_N\eta^\sh,H^{\fl N}],V]^\sh]\nn\\
  &+\frac12[[[[\xi,H^{\fl M}],\*_M\*_N\eta^\sh]^\sh,H^{\fl N}],V]\nn\\
  &+\frac12[[[[\*_M\*_N\eta^\sh,H^{\fl M}],\xi]^\sh,H^{\fl N}],V]-(\xi\leftrightarrow\eta)\;.
\end{align}
In the last term, we pull out the $H^{\fl M}$ at the price of a term
with $[\xi,H^{\fl M}]=\xi^Me_0$. Then this term cancels the first two
terms, and the remainder is
\begin{align}
  \label{RemainderTermII}
  -\frac12[[[[\xi,\*_M\*_N\eta^\sh],H^{\fl M}]^\sh,H^{\fl N}],V]-(\xi\leftrightarrow\eta)\;.
\end{align}

In the first step, the transformation parameter becomes
\begin{align}
    &-\frac12[[[\xi,\*_M\*_N\eta^\sh],H^{\fl M}]^\sh,H^{\fl N}]-(\xi\leftrightarrow\eta)\nn\\
    &=-\frac{(\lambda,\lambda)}{2((\lambda,\lambda)-1)}
    [[[\xi,\*_M\*_N\eta^\sh],H^{\fl M}],H^N]-(\xi\leftrightarrow\eta)\;,\nn\\
    &=-\frac{(\lambda,\lambda)}{2((\lambda,\lambda)-1)}
    [[[\xi,\*_M\*_N\eta^\sh],H^M],H^{\fl N}]-(\xi\leftrightarrow\eta)\;,
\end{align}
where the prefactor is to compensate for the factor in front of the
$T$ term in $[E,H]$, and where potential $\tk$ 
terms have been disregarded (they are easily shown to vanish).
The two right hand sides of this equation
can be calculated explicitly. In the Borcherds superalgebra, they give
the same result (there, the prefactor is absent).
In $S$, they give ``different'' expressions. The first one
is identical to the Borcherds algebra calculation: roughly speaking,
$[B^\fl_{MN},H^{\fl M}]$ gives only $R_1$, but the second one goes
through an intermediate $\tR_1$:
$[B^\fl_{MN},H^M]$ gives a combination of $R_1$ and $\tR_1$.
We use $\xi=\xi^ME_M$ etc. and calculate the expression contracting
$\xi^P\*_M\*_N\eta^Q$.
A straightforward calculation, using the section constraint,
gives the two alternative expressions
\begin{align}
  &-\frac{(\lambda,\lambda)}{2((\lambda,\lambda)-1)}\left(
  [[[E_P,H^{\fl M}],E^\sh_Q],H^N]+[[E_P,[E^\sh_Q,H^{\fl M}]],H^N]\right)\nn\\
  &\approx \frac12(2\delta_{[P}^Mt^\alpha_{Q]}{}^NT_\alpha
  -f^{\alpha\beta}{}_\gamma t^\alpha_P{}^Mt^\beta_Q{}^NT_\gamma)
\end{align}
and
\begin{align}
  &-\frac{(\lambda,\lambda)}{2((\lambda,\lambda)-1)}\left(
  [[[E_P,H^M],E^\sh_Q],H^{\fl N}]+[[E_P,[E^\sh_Q,H^M]],H^{\fl N}]\right)\nn\\
  &\approx\frac{(\lambda,\lambda)}{2((\lambda,\lambda)-1)}\left(
  2\delta_{[P}^Mt^\alpha_{Q]}{}^NT_\alpha
  -f^{\alpha\beta}{}_\gamma t^\alpha_P{}^Mt^\beta_Q{}^NT_\gamma\right)\nn\\
  &\qquad+\frac{1}{2((\lambda,\lambda)-1)}\ell_{\alpha P}{}^{\beta
    M}t^\alpha_Q{}^NT_\beta
\end{align}
respectively (the ``$\approx$'' sign denotes equality when the indices
are section-projected $\langle MN\rangle$ and antisymmetrised $[PQ]$). We recognise the $S$ tensor
in the first expression. The algebra now
identifies the result obtained by going through $R_1$ with the one
obtained by going through $\tR_1$, and the result becomes
\begin{align}
  -\frac12\ell_{\alpha P}{}^{\beta M}t^\alpha_Q{}^NT_\beta \;.
\end{align}
Unfortunately, the above calculation does not work for 
$(\lambda,\lambda)=1$, probably because the $q=0$ subalgebra then is
not $W(\fgr)$, but quite degenerate. This does not exclude that the
superalgebra $S(D_{r+1})$ provides a good description. There,
ancillary transformations are absent.

It is straightforward to show by explicit calculation
that the ancillary transformation
also can be expressed as
\begin{align}\label{XDefinition}
  \Sigma&=-\frac12[[[\xi^\sh,\*_M\*_N\eta^\sh],H^{\fl M}],H^{\fl N}]
  -(\xi\leftrightarrow\eta)\nn\\
&=-X^\fl_\xi\eta+X^\fl_\eta\xi\;.
\end{align}
The innermost bracket $[\xi^\sh,\*_M\*_N\eta^\sh]$ is in $\tR_2$, \ie,
a level $2$ element in $\fgrplus$. An ancillary element at ghost
number $1$ can be 
characterised as  $[[B_{MN},H^{\fl M}],H^{\fl N}]$,
where $MN$ are symmetric and in section (the antisymmetric part
vanishes due to the section constraint). Note, however, that its
appearance relies on a non-vanishing $\tR_1$.

The ancillary ghosts at ghost number $1$ are thus characterised as
doubly section-constrained objects constructed (through $\tR_1$) from
$\tR_2$. This is unlike higher ancillary ghosts $K_p$, $p\geq1$, which only need a
single section-constrained index, and are obtained as
$K_p^\fl=[B_M,H^\fl]$ with $B_M$ in $\tR_{p+1}$.

Let us consider the commutator of two ancillary transformations
$\Sigma$  and $\Sigma'$.
We write
\begin{align}
  \Sigma=-[[[\xi^\sh,\*_M\*_N\eta^\sh],H^{\fl M}],H^{\fl N}]
  =[[B_{MN},H^{\fl M}],H^{\fl N}]\;.
\end{align}
Let $V\in \tR_2$ at $(p,q)=(2,2)$ and let $s_{MNP}$ be a tensor whose
  all indices are in section. Then
$s_{MNP}[[[V,H^{\fl M}],H^{\fl N}],H^{\fl P}]=0$. This follows from
  the observation that $V$ contains irreducible representations
  $\overline{R(2\lambda-\alpha_i-\delta)}$ with $\delta$ in the
  positive root lattice. The consecutive commutators with $H^\fl$ contribute
  $R(3\lambda)$. The result must be in $R(\lambda)$, but this
  representation is not part of
  $R(3\lambda)\otimes\overline{R(2\lambda-\alpha_i-\delta)}$, since
$R(\lambda)\otimes R(2\lambda-\alpha_i-\delta)$ does not contain
  $R(3\lambda)$.
  If we then commute two ancillary generators, we immediately get
\begin{align}
  [\Sigma,\Sigma']=[[[\Sigma,B'_{MN}],H^{\fl M}],H^{\fl N}]
  =[[[B_{MN},\Sigma'],H^{\fl M}],H^{\fl N}]
\end{align}
which again is of the same type.

The same two-derivative form of the ancillary transformations is also
obtained by considering reducibility.
In the absence of ancillary transformations we had $\LL_{K^\fl}V=0$,
where $K$ is an ancillary ghost with ghost number $2$, obtained as
$K^\fl=[B_M,H^{\fl M}]$ with $B_M$ in $\tR_2$
(for higher ancillary ghosts the statement is
trivial).
Inserting this into the generalised
diffeomorphisms gives
\begin{align}
\LL_{K^\fl}V+[(dK)^\fl,V]=0\;,
\end{align}
so that now $K$ represents reducibility involving both the generalised
diffeomorphisms and the ancillary transformations.
Note that this consideration also gives ancillary ghosts at ghost
number $1$ constructed with two section-constrained indices from $\tR_2$.

\section{Dynamics\label{DynamicsSection}}

The remainder \eqref{RemainderTerm}
in the transformation of the part $\LL_0$ of the action contains the
tensor $S$ and arises only in situations when ancillary
transformations are present.
A candidate term, that is non-zero only in these cases, is suggested
by the ``new'' invariant tensor $\ell$, occurring as structure
constants in $S$ in eq. \eqref{EllTensor}, and projecting (with some weights) on the
irreducible modules in $\tR_1$:
\begin{align}
  \LL_1=\eta^{\alpha\gamma}G^{MP}
  \ell_{\alpha M}{}^{\beta N}\Pi_{P\beta}\Pi_{N\gamma}\;.
\end{align}
It has the right indices to contract a bilinear in $\Pi$ (together
with an inverse metric to match the weights) and vanishes in the
absence of ancillary transformations.
A straight-forward calculation gives at hand that the inhomogeneous
transformation cancels the one of $\LL_0$.
The calculation relies on the behaviour of $\ell$ under the
involution,
\begin{align}
  \ell_{\alpha M}{}^{\beta N}t^\alpha\otimes t_\beta
  =G_{MP}G^{NQ}\ell_{\beta Q}{}^{\alpha P}t^\star_\alpha\otimes t^{\star\beta}\;.
\end{align}
This property implies (thanks to
$\Pi_{M\alpha}t^\alpha=\Pi_{M\alpha}t^{\star\alpha}$) that the tensor
contracting the $\Pi$'s in $\LL_1$ is effectively symmetric under
$P\beta\leftrightarrow N\gamma$.
The inhomogeneous part of the variation becomes
\begin{align}
  \Delta_\xi\LL_1=2\ell_{\alpha M}{}^{\beta N}G^{MP}
  \left(t^{\alpha}{}_Q{}^R\Pi_P{}_\beta\*_N\*_R\xi^Q
  +     t_{\beta Q}{}^R\Pi_N{}^\alpha\*_P\*_R\xi^Q\right)\;.
\end{align}
The first term gives the $S$ tensor, thanks to the identity
\eqref{EllTIsS}, and precisely cancels $\Delta_\xi\LL_0$ in eq.
\eqref{RemainderTerm}.
In the second term, we need a ``new'' identity involving the section
condition in the indices $NR$.
Consider the invariant tensor occurring in the second term,
\begin{align}
  m_{\alpha M,Q}{}^{NR}\equiv\ell_{\alpha M}{}^{\beta\langle N}t_{\beta Q}{}^{R\rangle}
  +\ell_{\alpha M}{}^{\beta\{N}t_{\beta Q}{}^{R\}}\;.
\end{align}
The lower indices are in  sums of $R(\lambda)\otimes
R(\lambda-\beta)$, where again the $\beta$'s are lowest roots at level
$2$ or higher. This tensor product contains neither
$R(2\lambda)$ nor $R(2\lambda-\alpha_i)$ ($\lambda_i=1$), so $m$ is
identically $0$. The second term in the variation above vanishes.

The Lagrangian $\LL=\LL_0+\LL_1$ thus encodes
the complete dynamics for all extended geometries with
finite-dimensional structure group.

It is encouraging that the structure constants of the THA can be used
to construct an invariant Lagrangian. 
It seems quite clear that it will be possible to form the Lagrangian
as a combination of invariant contractions bilinear in the projections
of $dGG^{-1}$ on the torsion modules at level $-1$ in $W(\fgr)$. A task
in continued investigations will be to see if this specific
combination has a natural origin in the superalgebra.
However, $dGG^{-1}$ does not transform as a connection, but as the
symmetrised (with respect to the involution) part of a connection.
An alternative, but equivalent construction (see
ref. \cite{Cederwall:2015ica} for a discussion) is based on the
Weitzenb\"ock connection $dEE^{-1}$, where $E$ is a generalised
vielbein parametrising the coset $G/K$. The torsion part of this connection
does transform as a tensor, but then a specific combination of terms
will instead be dictated by invariance under local $K$
transformations, in complete analogy with the construction of the
action in the
teleparallel formulation of gravity.
It remains to be seen which is the most efficient way of formulating
the dynamics in terms of the superalgebra.

\section{$L_\infty$ algebra\label{LinftySection}}

The infinite tower of ghosts in exceptional field theory was first
described in ref. \cite{Berman:2012vc}, then without the introduction
of ancillary ghosts.
Ref. \cite{Palmkvist:2015dea} showed how the generalised
diffeomorphisms in exceptional field theory are constructed using a
Borcherds superalgebra, and this was generalised to the framework of
extended geometry in refs. \cite{Cederwall:2017fjm,Cederwall:2018aab}.
The $L_\infty$ algebra for double geometry was constructed in refs.
\cite{Deser:2016qkw,Hohm:2017pnh,Deser:2018oyg}.
In this case, there are no ancillary ghosts, and the algebra stops at
ghost number $2$ and a $3$-bracket. This corresponds to the Borcherds
superalgebra being finite-dimensional.
In ref. \cite{Cederwall:2018aab} the picture of
ref. \cite{Berman:2012vc} was refined by the introduction of ancillary
ghosts (with ghost number $>1$)
and the construction of the $L_\infty$ algebra (\ref{FullListBorcherds}), which completely
encodes the gauge structure of extended geometry in the absence of
ancillary gauge transformations.
We will now demonstrate that the THA
$S$ is the correct underlying algebraic structure in the
more general case. 

As discussed in Section \ref{THASection}, the THA $S$ essentially agrees
with the Borcherds
superalgebra $\BB(\fgrplus)$ at positive levels $p-q$.
A few differences arise that are relevant for the ghost structure, \ie, for the
$L_\infty$ algebra. The first is the presence of $\tR_1$, which makes
it possible to address the issue with ancillary ghosts at ghost number
$1$. The second is the presence of $T^\sh$ (but not $f_0$),
which is where these
ancillary ghosts actually reside.
Table \ref{THADerivativeActionTable} shows the generic structure of
the ghosts, with arrows showing the action of the nilpotent
``derivative'', the $L_\infty$ $1$-bracket.

\begin{table}
\begin{align*}
  \xymatrix{
    &&K_0\ar@{<-}[r]_d
    &K_1\ar[d]_\fl\ar@{<-}[r]_d
    &K_2\ar[d]_\fl\ar@{<-}[r]_d
    &K_3\ar[d]_\fl\ar@{<-}[r]_d&\cdots\\
    &&
    &C_1\ar@{<-}[r]_d
    &C_2\ar@{<-}[r]_d
    &C_3\ar@{<-}[r]_d&\cdots\\
}
\end{align*}
\caption{\it The typical structure of the action of the $1$-bracket
  between the ghost modules,
  with ancillary ghosts appearing from level $p=0$.}
\label{THADerivativeActionTable}
\end{table}

The only difference from the construction with the Borcherds
superalgebra in ref. \cite{Cederwall:2018aab} is that we now may have
ghosts in $K_0=k$. When formulating the $L_\infty$ brackets as derived
brackets based on $S$ instead of $\BB$, we can
to a large extent rely on the previous calculations, and ask how they
are modified by $k$.

The ghosts come in two kinds: the ``ordinary'' or non-ancillary ones
at $p>0$ and $q=0$, and the ancillary ones at $p\geq p_0$ and $q=1$, where
$\tR_{p_0+1}$ is the lowest occurrence of an $\tR_p$. In
$\BB$, we had $p_0\geq1$. For the cases presently under consideration,
with finite-dimensional $\fg$,
the corresponding limit in $S$ is $p_0\geq0$. (For
infinite-dimensional $\fg$, ancillary fields may appear at lower $p$,
see the discussion in Section \ref{DiscussionSection}.)
As for $\BB$, the non-ancillary ghosts are collectively denoted $C$ and the ancillary ones $K$,
but now seen as elements in $S$. The ancillary ghosts are
defined as in Section \ref{AncillarySection}.

We will not derive a full set of brackets and prove all identities,
but content ourselves with some low brackets, together with
conjectures on the general structure.

\subsection{Definitions and identities}
In ref. \cite{Cederwall:2018aab}, we needed to assume
the closure of the generalised diffeomorphisms, on the form
$\LL_C\LL_C=-\frac12\LL_{\LL_CC}$. The absence of ancillary
transformations had to be assumed, and did not follow from the content
of the Borcherds superalgebra --- even in situations with ancillary
transformations, there was no support for them in the
superalgebra. This prevented us from treating cases with ancillary
transformations. 

As in ref. \cite{Cederwall:2017fjm}, the generalised Lie derivative,
with one non-ancillary element $A$ as parameter, acting on another
non-ancillary element $B$, both of arbitrary statistics, was defined
as
\begin{align}
\LL_AB=\delta_{p_A,1}\left([[A,H^{\fl M}],\*_MB]
     +(-1)^{|B|}[[\*_MA^\sh,H^{\fl M}],B]\right)\;.
\end{align}
This expressions (now adapted to our notation for
the basis elements of $S$) still holds, since it does not involve ancillary
elements, and thus derives from isomorphic subalgebras of
$\BB$ and $S$.
Let $c=C_1\in R_1$ be non-ancillary ghost at ghost number $1$. 
In the presence of ancillary transformations, we have
(see eq. \eqref{XDefinition})
\begin{align}
  \label{LcLcA}
\LL_c\LL_cA=-\frac12\LL_{\LL_cc}A+\frac12[A,X^\fl_cc]\;,
\end{align}
where
\begin{align}
\label{XaB}
X^\fl_aB=-\frac12(-1)^{|B|}[[[\*_M\*_Na^\sh,B^\sh],H^{\fl M}],H^{\fl N}]\;,
\end{align}
for $a$ at $(p,q)=(1,0)$ and $B$ at $q=0$.

The $1$-bracket contains a horizontal part and a vertical part.
The vertical part is obtained from the lowering operator,
\begin{align}
A^\fl=-[A,e_0]\;.
\end{align}
The horizontal part is defined as 
\begin{align}
dA=[\*_MA^\sh,H^{\fl M}]
\end{align}
for any element $A$ such that $A^\fl=0$, \ie, for the lower element in
a pair.
It is then extended to the upper elements in the pairs (the ancillary
ghosts) by $(dK)^\fl=-dK^\fl$ for $K$ such that $K^\sh=0$. 
Then $d+\fl$ acts as a nilpotent $1$-bracket.
Also the generalised Lie derivative is extended to elements $K$ at height $1$ by
$(\LL_aK)^\fl=-\LL_aK^\fl$.

The generalised Lie derivative has the usual reducibility
$\LL_{dC_2}A=0$, where $C_2$ is an element at $(p,q)=(0,2)$. In
addition, there is a reducibility coming from parameters
$\dlb K_1\drb=dK_1+K_1^\fl$, where $K_1$ is an ancillary element at
$(p,q)=(1,1)$. The corresponding identity,
\begin{align}
\LL_{K_1^\fl}A-[A,(dK_1)^\fl]=0\;,
\end{align}
was derived in Section \ref{AncillarySection}.

The $2$-derivative expression $X_aB$ of eq. \eqref{XaB} appeared
already in the previous construction with the Borcherds superalgebra
\cite{Cederwall:2018aab}, however only for $B$ at $p\geq2$.
There it arose from the non-covariance of the derivative $d$ as
\begin{align}
  \label{dLLdX}
(d\LL_a+\LL_ad)B=-X^\fl_aB\;,
\end{align}
while we obtained it above from the commutator of two generalised
derivatives.
(For a discussion of the connection between non-covariance and the
appearance of ancillary transformations, see ref. \cite{Cederwall:2015ica}.)
We need to check that eq. \eqref{dLLdX} still holds when the
derivative acts on an ancillary ghost $K_1$, \ie, from $(p,q)=(1,1)$
to $(0,1)$. This is an issue, since
the corresponding (lowered) action from $(p,q)=(1,0)$ to $(0,0)$ is never
covariant, not even in the absence of ancillary transformations.
We have 
\begin{align}
(d\LL_a+\LL_ad)b=-X^\fl_ab+Y_ab\;,
\end{align}
where the non-ancillary contribution is
\begin{align}
Y_ab=-(-1)^{|b|}\*_M\*_Na^Pb^Nt^\alpha{}_M{}^PT_\alpha\;.
\end{align}
If $b$ is ancillary, \ie, $b=[\beta_M,H^{\fl M}]$ for $\beta_M$ in
  $\tR_2$, $Y_ab$ vanishes thanks to the antisymmetric section
  constraint.
The identity \eqref{dLLdX} can be used on all ghosts.

In ref. \cite{Cederwall:2018aab}, identities for commutators between
derivatives, generalised Lie derivatives and ancillary operators were
derived, and used in order to check the identities for the $L_\infty$
brackets.
We only need to consider modifications involving the presence of
$\tR_1$ and the ghost $k$.

The entities involved are the derivative $d$, the generalised Lie
derivative $\LL_aB$, the two-derivative ancillary element $X_aB$ of
eq. \eqref{XaB} and the one-derivative ancillary element $R(A,B)$
defined by
\begin{align}
R^\fl(A,B)=(-1)^{|B|}\frac{p_B\*_M^{(A)}-p_A\*_M^{(B)}}{p_A+p_B}[[A^\sh,B^\sh],H^{\fl M}]^\fl\;.
\end{align}
The latter expression roughly indicates the deviation of $d$ from
being a derivation. Let $a,b$ and $A,B$ be elements at $q=0$ and let $p_a=p_b=1$,
$p_A,p_B>1$. Then,
\begin{align}
d[a,b]&=\LL_ab-(-1)^{|a||b|}\LL_ba-R^\fl(a,b)\;,\nn\\
d[a,B]&=[a,dB]+\LL_aB-R^\fl(a,B)\;,\label{DerivationDeviation}\nn\\
d[A,B]&=[A,dB]+(-1)^{|B|}[dA,B]-R^\fl(A,B)\;.
\end{align} 
In addition it is straightforward to derive
\begin{align}
d[a,\alpha^\fl]=\LL_a\alpha^\fl-R^\fl(a,\alpha^\fl)\;,\label{daalpha}
\end{align}
where $a$ is an element at $(p,q)=(1,0)$ and $\alpha$ is an ancillary
element in the adjoint at $(p,q)=(0,1)$.

We will not give a full list of identities. Most of them, except for
very low ghost number, coincide with the ones in
ref. \cite{Cederwall:2018aab}. 
From the expression \eqref{dLLdX} it immediately follows that
\begin{align}
(dX_a+X_ad)B=0\;.
\end{align}
The explicit expression for $R(a,b)$ gives
\begin{align}
  \label{dRabXab}
dR(a,b)=X_ab-(-1)^{|a||b|}X_ba\;.
\end{align}
Another identity at $(p,q)=(0,1)$ 
is
\begin{align}
  \label{LXCommutator}
\LL_cX_cc+X_c\LL_cc=-\frac12X_{\LL_cc}c+\frac12R(c,X^\fl_cc)
\end{align}
which can be proven by comparing the explicit expressions. It
generalised the corresponding relation
``$\LL_cX_cA+X_c\LL_cA=-\frac12X_{\LL_cc}A$'' in the Borcherds case.

\subsection{Some $L_\infty$ brackets}
When we truncate to the ghost sector, we in addition postulate that
the $1$-bracket annihilates the lowest ghosts.
The $1$-bracket is
\begin{align}
\dlb c\drb&=0\;,\nn\\
\dlb C_p\drb&=dC_p\;, \quad p\geq2\;,\nn\\
\dlb k\drb&=0\;,\nn\\
\dlb K_p\drb&=dK_p+K_p^\fl\;,\quad p\geq1\;,
\end{align}
where we denote the ghost number $1$ non-ancillary ghost $c=C_1$ and
the ancillary one $k=K_0$ (the subscript is the $p$ eigenvalue, not
the ghost number).
This is depicted in Table \ref{THADerivativeActionTable}.

Let us start with the $2$-brackets between elements at ghost number
$1$. We let $c\in R_{(1,0)}$  and $k\in R_{(0,1)}$. The
$1$-bracket by definition annihilates $c+k$, $\dlb c+k\drb=0$.
The $2$-brackets reflect the commutators between transformations, and
we have, using the form of the ancillary transformations derived in
Section \ref{AncillarySection},
\begin{align}
  \dlb c,c\drb&=\LL_cc+X_cc\;,\nn\\
  \dlb c,k\drb_0&=\LL_ck\;,\label{ckTwoBrackets}\nn\\
  \dlb k,k\drb&=-[k,k^\fl]
\end{align}
The $2$-bracket identities are of course trivially satisfied.
The second of these brackets have been equipped with subscript $0$,
since it will be modified.
When the ghost $k$ is present, the bracket $\dlb c,k\drb$ is not
uniquely determined by the commutator of a generalised Lie derivative and an
ancillary transformation. The action of the commutator on \eg\ a
vector is of course unique, but the corresponding set of parameters is
not. We can choose to add a trivial term proportional to
\begin{align}
  \label{TrivialParameters}
  \dlb[c,k]^{\fl\sh}\drb=-(d[c,k^\fl])^\sh+[c,k^\fl]
  =\LL_ck+R(c,k^\fl)+[c,k^\fl]\;,
\end{align}
where we have used eq. \eqref{daalpha} in the last step,
representing a vanishing tranformation due to reducibility.
The operation $\fl\sh$ ensures that the parameter lies in $R_1$ at
height $1$, and not in $\tR_1$.
The choice made here will have repercussions for the brackets
containing higher ghost number ghosts and for higher brackets.
A goal is to connect as closely as possible
to the results of ref. \cite{Cederwall:2018aab},
where choices for low brackets enabled us to give universal
expressions for all brackets between all ghosts.
In particular, we then had $\dlb C,K\drb=\frac12\LL_CK$, with a factor
which contrasts with the middle equation in \eqref{ckTwoBrackets}.
It will serve our purposes to add the trivial transformation
of eq. \eqref{TrivialParameters} with coefficient $-\frac12$ to obtain
\begin{align}
  \dlb c,c\drb&=\LL_cc+X_cc\;,\nn\\
  \dlb c,k\drb&=\LL_ck+\frac12\dlb[c,k]^{\fl\sh}\drb
  =\frac12\LL_ck-\frac12R(c,k^\fl)-\frac12[c,k^\fl]\;,\nn\\
  \dlb k,k\drb&=-[k,k^\fl]\label{ckTwoBracketsMod}
\end{align}

The ``algebra'' of eq. \eqref{ckTwoBrackets}
is a simple choice of representative
in that it
contains no generalised diffeomorphisms in the commutator
between a generalised diffeomorphism and an ancillary
transformation. The representative in eq. \eqref{ckTwoBracketsMod}, on
the other hand, is not of this kind, but has other advantages.

In ref. \cite{Cederwall:2018aab}, the coefficient of the $n$-brackets
contained the Bernoulli number $B_{n-1}$. The vanishing of the
Bernoulli numbers for odd argument $\geq3$ implied the vanishing of
the even brackets, starting from the $4$-bracket.
It would be desirable to maintain this property. This turns out to be
possible, using the choice above and further choices for higher
brackets.

Before completing the $2$-brackets, we consider the $3$-bracket
$\dlb c,c,c\drb$. The identity is
\begin{align}
\dlb\dlb c,c,c\drb\drb+2\dlb c,\dlb c,c\drb\drb=0\;.
\end{align}
The second term is calculated as
\begin{align}
2\dlb c,\dlb c,c\drb\drb&=2\dlb c,\LL_cc+X_cc\drb\nn\\
&=\LL_c\LL_cc+\LL_{\LL_cc}c\nn\\
&\quad+X_c\LL_cc+X_{\LL_cc}c+\LL_cX_cc
-R(c,X^\fl_cc)-[c,X^\fl_cc]
\;.
\end{align}
The terms at $q=0$ (the first two terms, together with the last one,
coming from the modification of the bracket $\dlb c,k\drb$)
can be rewritten using the
first equation in \eqref{DerivationDeviation} together with
eq. \eqref{LcLcA} as
\begin{align}
  \LL_c\LL_cc+\LL_{\LL_cc}c-[c,X^\fl_cc]
  =-\frac13\left(d[c,\LL_cc]+R^\fl(c,\LL_cc)+[c,X^\fl_cc]\right)\;.
\end{align}
This immediately gives the full $3$-bracket as
\begin{align}
\dlb c,c,c\drb=\frac13\Bigl\{[c,\LL_cc]+R(c,\LL_cc)+[c,X_cc]\Bigr\}
\end{align}
It remains to be verified that the $q=1$ part of the identity is
satisfied.
This demands that
\begin{align}
  \label{cccIdentity}
  \frac13dR(c,\LL_cc)+\frac13d([c,X_cc]^{\fl\sh})+\LL_cX_cc+X_c\LL_cc+X_{\LL_cc}c
  -R(c,X^\fl_cc)=0\;.
\end{align}
A short calculation, using eqs. \eqref{dRabXab} and \eqref{daalpha},
shows the the $q=1$ part of the identity holds, thanks to eq.
\eqref{LXCommutator}.
Note that $3$-bracket has the same formal
expression as in eq. \eqref{FullListBorcherds}. In particular, the
coefficients of $[c,\LL_cc]$ and $[c,X_cc]$ are the same,
and this happens only for our
particular choice of representative for $\dlb c,k\drb$.

Let us now address the vanishing of the $4$-bracket $\dlb
c,c,c,c\drb$.
The identity to be fulfilled is
\begin{align}
  \dlb\dlb c,c,c,c\drb\drb+2\dlb c,\dlb c,c,c\drb\drb
  +3\dlb c,c,\dlb c,c\drb\drb=0\;,
\end{align}
and we will for now only consider the $q=0$ part.
Assuming that $\dlb c,K_1\drb$ does not contain a $q=0$ part (see below), it
can be calculated as
\begin{align}
  &\dlb\dlb c,c,c,c\drb\drb\vert_{q=0}
  +\frac23[c,\LL_c\LL_cc+\frac12\LL_{\LL_cc}c]+3\dlb
    c,c,X_cc\drb\vert_{q=0}\nn\\
    &=\dlb\dlb c,c,c,c\drb\drb\vert_{q=0}
    +\frac13[c,[c,X^\fl_cc]]+3\dlb c,c,X_cc\drb\vert_{q=0}
\;.  
\end{align}
A necessary condition for the consistency of the vanishing of the
$4$-bracket is that $\dlb c,c,k\drb$ contains a $q=0$ part
$-\frac19[c,[c,k^\fl]]$. This can be arranged by choosing a
representative by the suitable addition of
a trivial term proportional to
$\dlb [c,[c,k]]^{\fl\sh}\drb$. It also becomes clear --- by the
calculations in the Borcherds case together with the ancillary term in
$\LL_c\LL_c$ in eq. \eqref{LcLcA} --- that
this happens to all orders, and that we can reproduce the
collective brackets
\begin{align}
\dlb C^n\drb=k_n\Bigl((\ad C)^{n-2}(\LL_CC+X_CC)
+\sum\limits_{i=0}^{n-3}(\ad C)^iR_C(\ad C)^{n-i-3}\LL_CC\Bigr)\;,
\end{align}
as before, with the new extended meaning of $X_CC$.

Let us now consider brackets with one ancillary ghost.
The lowest identity is
\begin{align}
  0&=\dlb\dlb c,K_1\drb\drb+\dlb c,\dlb K_1\drb\drb\nn\\
  &=\dlb\dlb c,K_1\drb\drb+\dlb c,dK_1\drb+\dlb c,K_1^\fl\drb\;.
\end{align}
The last two terms become, using  eq. \ref{ckTwoBracketsMod},
\begin{align}
  \dlb c,dK_1\drb+\dlb c,K_1^\fl\drb
  &=\LL_cdK_1-\frac12R(c,dK_1)-\frac12[c,(dK_1)^\fl]\nn\\
  &\quad\,+\frac12\LL_cK_1^\fl+\frac12\LL_{K_1^\fl}c
  +\frac12X_cK_1^\fl+\frac12X_{K_1^\fl}c\;.
\end{align}
Using the reducibility $\LL_{K_1^\fl}c-[c,(dK_1)^\fl]=0$, the height
$0$ terms become $-\frac12\LL_cK_1$, implying that
\begin{align}
\dlb c,K_1\drb=\frac12\LL_cK_1\,,
\end{align}
which agrees with the expression in the Borcherds case.
The remaining terms at height $1$ of the identity demand are
\begin{align}
  \frac12\left(d\LL_cK_1+\LL_cdK_1
  +X_cK_1^\fl+X_{K_1^\fl}c-R(c,(dK_1)^\fl\right)&\nn\\
  =\frac12\left(X_{K_1^\fl}c-R(c,(dK_1)^\fl)\right)&=0\;.
\end{align}
This result is readily extended to
\begin{align}
\dlb C,K'\drb=\frac12\LL_CK'\;,
\end{align}
where $K'=K-k$ is the $p\geq1$ part of $K$, in accordance with the
Borcherds case.

For the $2$-brackets with two ancillary ghosts one easily obtains
\begin{align}
  \dlb k,K'\drb&=-\frac12[K',k^\fl]\;,\nn\\
  \dlb K',K'\drb&=0\;.
\end{align}

This completes the $2$-brackets between all ghosts.
It is possible to introduce the notation
$\MM_KA=-[A,k^\fl]$ and $Y_KA=-R(A,k^\fl)$ (in analogy with
$\LL_CA=\LL_cA$), in order to write
the $2$-brackets collectively as
\begin{align}
  \dlb C,C\drb&=\LL_CC+X_CC\;,\nn\\
  \dlb C,K\drb&=\frac12(\LL_CK+\MM_KC+Y_KC)\;,\nn\\
  \dlb K,K\drb&=\MM_KK\;.
\end{align}

It is clear that the construction works, and that there are solutions
to the identities that make the higher brackets take forms
close to the ones in the Borcherds case. We conjecture that all
brackets reduce to the formal expressions for the Borcherds ones
under $k\rightarrow0$, that all brackets with more than two ancillary
ghosts vanish, and that all even brackets above the $2$-bracket vanish.

\section{Discussion}\label{DiscussionSection}

We have given a description of the dynamics and gauge structure of extended
geometry with a finite-dimensional structure group $G$ and lowest
weight coordinate representation $R(-\lambda)$. The tensor hierarchy
algebra
$S=S(\fgrplus)$ plays a central r\^ole, in that it
(unlike the corresponding Borcherds superalgebra)
naturally
harbours the ancillary transformations.
Ingredients from the THA were also used for a
description of the dynamics.

One main purpose of the paper, and its companion
\cite{CederwallPalmkvistTHAI}, is to find the underlying algebraic
structure behind extended geometry. The THA $S$,
in every respect, shows promise to
 contain exactly the correct information
precisely when it is needed. One of the relevant aspects for the
present investigation is the peculiar appearance of new elements
(in the present case, a lowest weight $\fgrplus$-module, starting with
the generators $L_{\alpha M}$) along
with $\fgrplus$ at level $0$ (\cf\ ref. \cite{Bossard:2017wxl}).
It is tempting to extrapolate to more complicated situations,
especially with infinite-dimensional structure groups
\cite{Bossard:2017aae}, and follow the lead given by the THA.
For $G=E_9$, the relevant THA
algebra is $S(E_{10})$.
The central part of this superalgebra (or more generally, for any
affine $G$), decomposed into $E_9$
representations along the same principles as previous tables, is
given in Table \ref{HyperbolicSTableBasis}.
In order to construct this part of the algebra, one needs to include
the generator $\dd$ (the Virasoro generator $L_0$, roughly speaking)
in order to have a non-degenerate Cartan matrix (bilinear form).

\begin{table}
  \begin{align*}
  \xymatrix@=.4cm{
    \ar@{-}[]+<2em,1em>;[dddd]+<2em,-1em>
    \ar@{-}[]+<-0.8cm,-1em>;[rrr]+<1.4cm,-1em>
    &\ar@{-}[]+<2.65em,1em>;[dddd]+<2.65em,-1em> p=-1
    & \ar@{-}[]+<3.5em,1em>;[dddd]+<3.5em,-1em> p=0 &p=1\\
q=2&&\shift{1}{\pi^\sh}&\shift{1}{L^\sh_M}\\
q=1&\shift{0}{\Phi^{\sh M}}&\shift{1}{T^{\sh A}_m}\;\shift{1}{\KK^\sh}\;
       \shift{0}{\pi}\;\shift{0}{L^\sh_1}
   & \shift{1}{E_M^\sh}\;\shift{0}{L_M}
       \\ 
q=0&\shift{-1}{\Phi^M}\;\shift{0}{H^M}
&\shift{0}{T^A_m}\;\shift{0}{\KK}\;\shift{0}{\dd}\;\shift{-1}{L_1}
           & \shift{0}{E_M}\\ 
q=-1 & \shift{-1}{H^{\fl M}}& \shift{-1}{e_0} 
  }
\end{align*}
\vspace{-.5cm}
  \caption{\it Basis elements for $S(\fgrplus)$ when $\fgr$ is
    an affine algebra.
    The shifts specifying the action of $\dd$ are given in red.}
\label{HyperbolicSTableBasis}
\end{table}

Here, the generators at $(p,q)=(0,0)$ are the affine generators,
including $\dd$, which is $L_0$, but acting with a ``shift'' compared
to $L_0$, indicated in red below the generators. There is also $L_1$,
which we have already encountered in $S(E_9)$. The elements at
$p=-1$ and $p=1$  are (shifted) fundamentals and
anti-fundamentals of $E_9$.
It is noteworthy that now even $\tR_0$ is non-empty, and consists of a
singlet. We have seen that the presence of $\tR_p$ indicates ancillary
ghosts at ghost number $p$, compensating for an apparent failure in the
$(p+1)$-bracket.
It will be interesting to see what this means for the $1$-bracket in 
$E_9$ geometry.
In the $E_9$ case, or more generally in the affine case, including the
geometrisation of the Geroch group, this seems to point at the
appearance of ancillary fields already at ghost number $0$,
represented by constrained fields in the unshifted fundamental
representation at $(p,q)=(-1,1)$. This would corroborate and provide
an algebraic basis for the results of ref. \cite{Bossard:2018utw},
and will be the subject of a forthcoming study.

The main purpose of the invention of THA's
\cite{Palmkvist:2013vya} was to have the embedding tensor at level
$-1$. This representation ($\Phi$) together with a ``fundamental''
($H$) form the representations in which torsion comes in extended
geometry. It is likely that the ghost sector can be complemented with
dynamical fields at level $0$, ``torsion''
antifields at level $-1$, and further higher antifields corresponding
to torsion Bianchi identities etc. (``syzygies'') in the algebraic
framework. This is of course completely beyond the reach of a
Borcherds superalgebra.

The fact that the generalised Lie derivative fulfils a Leibniz rule means that the vector fields satisfy a Leibniz algebra
\cite{Hohm:2018ybo,Hohm:2019wql}. It has recently been shown that any Leibniz algebra canonically (via a differential graded
Lie algebra, or equivalently, an infinity-enhanced Leibniz algebra \cite{Bonezzi:2019ygf}) gives rise to an $L_\infty$ algebra \cite{Lavau:2019oja}.
It would be interesting to compare the $L_\infty$ algebra constructed in that way with the one presented here,
not least since 
in the application to gauged supergravity, 
the relevant differential graded
Lie algebra can be understood as coming from a tensor hierarchy algebra \cite{Greitz:2013pua,Lavau:2017tvi}.

In some cases, notably in the exceptional series, the THA's
of $S$ type possess a non-degenerate invariant bilinear form
\cite{Palmkvist:2013vya,Bossard:2017wxl}, which is invariant under the
algebra but not centered around level $0$. For $S(E_{r+1})$, a
$\fgr$-module at $(p,q)$ in the double grading is paired with the
conjugate module at $(9-r-p,1-q)$.
Given an involution on $\fgr$,
one may use it to define a ``dualisation'' map
$\star\in\End(S(\fgrplus))$ and, as proposed in \cite{Bossard:2017wxl},
the bilinear form might be used in
an action.

\section*{Acknowledgements}
MC wants to thank E. Almqvist for discussions on factors of $\frac12$.
This research is supported by the
Swedish Research Council, project no. 2015-04268.

\appendix

\bibliographystyle{utphysmod2}


\providecommand{\href}[2]{#2}\begingroup\raggedright\endgroup

\end{document}